\documentclass{easychair}

\usepackage{makeidx}

\makeindex

\newcommand{\xs}[1]{Section~\ref{#1}}

\newcommand{\xt}[1]{Table~\ref{#1}}

\newcommand{\xe}[1]{Equation~\ref{#1}}

\newcommand{\C}{{C\index{C}}}
\newcommand{\cpp}{{C++\index{C++}}}

\newcommand{\java}{{Java\index{Java}}}

\newcommand{\file}[1]{\url{#1}\index{Files!#1}}
\newcommand{\tool}[1]{\texttt{#1}\index{Tools!#1}}
\newcommand{\option}[1]{\texttt{#1}\index{Options!#1}}
\newcommand{\api}[1]{\texttt{#1}\index{API!#1}}

\newcommand{\marf}[0]{MARF\index{MARF}\index{Frameworks!MARF}\index{Libraries!MARF}}

\newcommand{\lucidL}[1]{{$\mathit{Lucid}$}($L$) }

		{}

\def\myvert{\raise 2.27pt \hbox{\vrule depth 0pt height 8pt width 0.2mm}}
\def\myarrow{\hspace*{0.43mm}%
             \raise 2.29pt\hbox{\vrule depth 0pt height 8pt width 0.16mm}%
             \hspace*{-0.32mm}%
             $\longrightarrow$
             \ %
             }

\newcommand{\ngram}{$n$-gram}
\newcommand{\marfcat}{MARFCAT\index{MARFCAT}\index{MARF!Applications!MARFCAT}}

\newcommand
	{\cve}
	[1]
	{\href{http://cve.mitre.org/cgi-bin/cvename.cgi?name=#1}{#1}\index{CVE!#1}}

\newcommand
	{\cwe}
	[1]
	{#1\index{CWE!#1}}

\newcommand
	{\chromeCase}
	[1]
	{Chrome 5.0.375.#1\index{Chrome!5.0.375.#1}\index{Test cases!Chome 5.0.375.#1}}

\newcommand
	{\wiresharkCase}
	[1]
	{Wireshark 1.2.#1\index{Wireshark!1.2.#1}\index{Test cases!Wireshark 1.2.#1}}

\newcommand
	{\dovecotCase}
	{Dovecot\index{Dovecot}\index{Test cases!Dovecot 2.0.beta6.20100626}}

\newcommand
	{\tomcatCase}
	[1]
	{Tomcat 5.5.#1\index{Tomcat!5.5.#1}\index{Test cases!Tomcat 5.5.#1}}

\newcommand
	{\pebbleCase}
	{Pebble\index{Pebble}\index{Test cases!Pebble 2.5-M2}}

\newcommand{\NIST}{}

\begin{document}

\title{The use of machine learning with signal- and NLP processing of
source code to fingerprint, detect, and classify vulnerabilities and weaknesses
with MARFCAT}

\titlerunning{MARFCAT: A MARF Approach to SATE2010}

\author{Serguei A. Mokhov\\
\affiliation{Concordia University}\\
\affiliation{Montreal, QC, Canada}\\
\affiliation{\url{mokhov@cse.concordia.ca}}}

\authorrunning{S. A. Mokhov}

\maketitle

\NIST{}
\NIST{}

\begin{abstract}
We present a machine learning approach to static code analysis and findgerprinting
for weaknesses related to security, software engineering, and others
using the open-source {\marf} framework and the {\marfcat} application
based on it for the NIST's SATE 2010 static analysis tool exposition
workshop.
\end{abstract}

\tableofcontents
\listoftables

\section{Introduction}
\label{sect:introduction}

This paper elaborates on the details of the methodology
and the corresponding results of application of the machine
learning techniques along with signal processing and NLP alike
to static source code analysis in search for weaknesses and
vulnerabilities in such a code. This work resulted in a proof-of-concept tool,
code-named {\em {\marfcat}}, a {\marf}-based Code Analysis Tool \cite{marfcat-app}, presented
at the Static Analysis Tool Exposition (SATE) workshop 2010 \cite{nist-samate-sate2010} collocated
with the Software Assurance Forum on October 1, 2010.

This paper is a ``rolling draft'' with several updates expected
to be made before it reaches more complete final-like version
as well as combined with the open-source release of the MARFCAT tool
itself \cite{marfcat-app}.
As-is it may contain inaccuracies and incomplete information.

At the core of the workshop there were {\C}/{\cpp}-language and {\java}
language tracks comprising CVE-selected cases as well as stand-alone cases.
The CVE-selected cases had a vulnerable version of a software in question
with a list of CVEs attached to it, as well as the most know fixed version
within the minor revision number. One of the goals for the CVE-based cases
is to detect the known weaknesses outlined in CVEs using static code analysis
and also to verify if they were really fixed in the ``fixed version'' \cite{nist-samate-sate2010}.

\noindent
The test cases at the time included CVE-selected:

\begin{itemize}
	\item {\C}: \wiresharkCase{0} (vulnerable) and \wiresharkCase{9} (fixed)
	\item {\cpp}: \chromeCase{54} (vulnerable) and \chromeCase{70} (fixed)
	\item {\java}: \tomcatCase{13} (vulnerable) and \tomcatCase{29} (fixed)
\end{itemize}

\noindent
and non-CVE selected:

\begin{itemize}
	\item {\C}: {\dovecotCase} 2.0-beta6
	\item {\java}: {\pebbleCase} 2.5-M2
\end{itemize}

We develop {\marfcat} to machine-learn from the CVE-based vulnerable cases
and verify the fixed versions as well as non-CVE based cases from similar
programming languages.

\subsection*{Organization}

We develop this ``running'' article gradually.
The related work, some of the present methodology is based on, is referenced in \xs{sect:related-work}.
The methodology summary is in \xs{sect:methodology}.
We present the results, most of which were reported at SATE2010, in \xs{sect:results}.
We then describe the machine learning aspects as well as mathematical estimates of functions
of how to determine line numbers of unknown potentially weak code fragments in \xs{sect:line-numbers}.
(The latter is necessary since during the representation of the code a wave form (i.e. signal)
with current processing techniques the line information is lost (e.g. filtered out as noise)
making reports less informative, so we either machine-learn the line numbers or provide
a mathematical estimate and that section describes the proposed methodology to do so,
some of which was implemented.) Then we present a brief summary,
description of the limitations of the current realization of the approach and
concluding remarks in \xs{sect:conclusion}.

\section{Related Work}
\label{sect:related-work}

Related work (to various degree of relevance)
can be found below (this list is not exhaustive):

\begin{itemize}
	\item 
Taxonomy of Linux kernel vulnerability solutions in terms of
patches and source code as well as categories for both are found in
\cite{linux-vuln-sols-cisse07}.
	\item 
The core ideas and principles behind the {\marf}'s pipeline and testing
methodology for various algorithms in the pipeline adapted to this case
are found in \cite{marf-c3s2e08}. There also one can find the core options
used to set the configuration for the pipeline in terms of algorithms used.
	\item 
A binary analysis using machine learning approach for quick scans for
files of known types in a large collection of files is described in
\cite{marf-file-type}.
	\item 
The primary approach here is similar in a way that was done for
DEFT2010 \cite{marf-deft,marf-deft-complete-results} with the
corresponding \api{DEFT2010App} and its predecessor
\api{WriterIdentApp} \cite{marf-writer-ident}.
	\item 
Tlili's 2009 PhD thesis covers topics on automatic detection of safety and security
vulnerabilities in open source software \cite{tlili-phd-vuln-detect-oss-2009}.
	\item 
Statistical analysis, ranking, approximation, dealing with uncertainty,
and specification inference in static code analysis are found in the
works of Engler's team \cite{stats-spec-within,correlation-error-ranking,z-ranking-2003}.
	\item 
Kong et al. further advance static analysis (using parsing, etc.) and specifications
to eliminate human specification from the static code analysis in 
\cite{no-human-spec-static-analysis}.
	\item
Spectral techniques are used for pattern scanning in malware detection by Eto et al. in
\cite{malware-spectrum-analysis-2009}.
	\item
Researchers propose a general data mining system for incident analysis with data mining
engines in \cite{incident-analysis-nicter-data-mining-wistdcs-2008}.
	\item
Hanna et al. describe a synergy between static and dynamic analysis for the detection
of software security vulnerabilities in \cite{synergy-static-dynamic-2009} paving the way
to unify the two analysis methods.
	\item
The researchers propose a MEDUSA system for metamorphic malware dynamic analysis
using API signatures in \cite{medusa-malanal-api-sin-2010}.
\end{itemize}

\section{Methodology}
\label{sect:methodology}

Here we briefly outline the methodology of our approach to
static source code analysis in its core principles in 
\xs{sect:core-principles}, the knowledge base in \xs{sect:kb},
machine learning categories in \xs{sect:ml-cats}, and
the high-level step-wise description in \xs{sect:basic-methodology}.

\subsection{Core principles}
\label{sect:core-principles}

The core methodology principles include:

\begin{itemize}
\item Machine learning
\item Spectral and NLP techniques
\end{itemize}

We use signal processing techniques, i.e. presently we do not parse
or otherwise work at the syntax and semantics levels. We treat the source code as a ``signal'',
equivalent to binary, where each {\ngram} ($n=2$ presently, i.e. two  
consecutive characters or, more generally, bytes) are used to  
construct a sample amplitude value in the signal.

We show the system examples of files with weaknesses and {\marfcat} learns them
by computing spectral signatures using signal processing techniques from CVE-selected test cases.
When some of the mentioned techniques are applied (e.g. filters, silence/noise removal, other
preprocessing and feature extraction techniques), the line number information
is lost as a part of this process.

When we test, we compute how similar
or distant each file is from the known trained-on weakness-laden files.
In part, the methodology can approximately be seen as some signature-based antivirus or IDS software systems
detect bad signature, except that with a large number of machine learning and signal processing
algorithms, we test to find out which combination gives the highest
precision and best run-time.

At the present, however, we are looking at the files overall
instead of parsing the fine-grained details of patches and weak code fragments,
which lowers the precision, but is fast to scan all the files.

\subsection{CVEs -- the ``Knowledge Base''}
\label{sect:kb}

The CVE-selected test cases serve as a source of the knowledge
base to gather information of how known weak code ``looks like''
in the signal form, which we store as spectral signatures
clustered per CVE or CWE. Thus, we:

\begin{itemize}
\item Teach the system from the CVE-based cases 
\item Test on the CVE-based cases
\item Test on the non-CVE-based cases
\end{itemize}

\subsection{Categories for Machine Learning}
\label{sect:ml-cats}

The tow primary groups of classes we train and test on
include:

\begin{itemize}
\item CVEs \cite{nist,niststats}
\item CWEs \cite{mitre-cwes} and/or our custom-made, e.g. per our classification methodology in \cite{linux-vuln-sols-cisse07}
\end{itemize}

The advantages of CVEs is the precision and the associated
meta knowledge from \cite{nist,niststats} can be all aggregated
and used to scan successive versions of the the same software
or derived products. CVEs are also generally uniquely mapped
to CWEs.
The CWEs as a primary class, however, offer broader categories,
of kinds of weaknesses there may be, but are not yet well
assigned and associated with CVEs, so we observe the loss of
precision.

Since we do not parse, we generally cannot deduce weakness types
or even simple-looking aspects like line numbers where the weak
code may be. So we resort to the secondary categories, that are
usually tied into the first two, which we also machine-learn
along, shown below:

\begin{itemize}
\item Types ({\em sink}, {\em path}, {\em fix})
\item Line numbers
\end{itemize}

\subsection{Basic Methodology}
\label{sect:basic-methodology}

Algorithmically-speaking, {\marfcat} performs the following steps to do its
learning analysis:

\begin{enumerate}
\item
Compile meta-XML files from the CVE reports (line numbers, CVE, CWE, fragment size, etc.). Partly done by a Perl script and partly manually.
This becomes an index mapping CVEs to files and locations within files.
\item
Train the system based on the meta files to build the knowledge base (learn).
Presently in these experiments we use simple mean clusters of feature vectors
per default {\marf} specification (\cite{marf-c3s2e08,marf}).
\item
Test on the training data for the same case (e.g. \tomcatCase{13} on \tomcatCase{13}) with the same annotations
to make sure the results make sense by being high and deduce the best algorithm combinations for the task.
\item
Test on the testing data for the same case (e.g. \tomcatCase{13} on \tomcatCase{13}) without the annotations
as a sanity check.
\item
Test on the testing data for the fixed case of the same software (e.g. \tomcatCase{13} on \tomcatCase{29}).
\item
Test on the testing data for the general non-CVE case (e.g. \tomcatCase{13} on {\pebbleCase}).
\end{enumerate}

\subsection{Line Numbers}
\label{sect:line-numbers}

As was earlier mentioned, line number reporting with {\marfcat} is an issue
because the source text is essentially lost without line information preserved
(filtered out as noise or silence or mixed in with another signal sample).
Therefore, some conceptual ideas were put forward to either derive a heuristic, a function
of a line number based on typical file attributes as described below, or learn
the line numbers as a part of the machine learning process. While the methodology
of the line numbers discussed more complete scenarios and examples, only
and approximation subset was actually implemented in {\marfcat}.

\subsubsection{Line Number Estimation Methodology}

Line number is a function of the file's dimensions in terms
of line numbers, size in bytes, and words. The meaning of
$W$ may vary. The implementations of $f$ may vary and can be
purely mathematical or relativistic and with side effects.
These dimensions were recorded in the meta XML files along
with the other indexing information. This gives as the
basic \xe{eq:line-f-basic}.

\begin{equation}
l = f(L_T, B, W)
\label{eq:line-f-basic}
\end{equation}

\noindent
where
\begin{itemize}
	\item $L_T$ -- number of lines of text in a file
	\item $B$ -- the size of the file in bytes
	\item $W$ -- number of {\em words} per \tool{wc} \cite{wc}, but can be any blank delimited
	printable character sequence; can also be an {\ngram} of $n$ characters.
\end{itemize}

\noindent
The function should be additive to allow certain components
to be zero if the information is not available or not needed,
in particular $f(B)$ and $f(W)$ may fall into this category.
The ceiling $\lceil \ldots \rceil$ is required when functions
return fractions, as shown in \xe{eq:line-f-basic-def}.

\begin{equation}
f(L_T, B, W) = \lceil f(L_T) + f(B) + f(W) \rceil
\label{eq:line-f-basic-def}
\end{equation}

\noindent
Constraints on parameters:

\begin{itemize}
	\item $l \in [1,\ldots,L_T]$ -- the line number must be somewhere within the lines of text.
	\item $f(L_T) > 0$ -- the component dependent on the the lines of text $L_T$ should never be zero or less.
	\item $EOL = \{\mathtt{\string\n}, \mathtt{\string\r}, \mathtt{\string\r\string\n}, \mathtt{EOF}\}$. The inclusion
	of \texttt{EOF} accounts for the last line of text missing the traditional line endings, but is non-zero.
	\item $L_T > 0 \implies B > 0$
	\item $B > 0 \implies L_T > 0$ under the above definition of \texttt{EOL}; if \texttt{EOF} is excluded this implication would not be true
	\item $B = 0 \implies L_T=0, W=0$
\end{itemize}

\noindent
Affine combination is in \xe{eq:line-f-def-affine}:

\begin{equation}
f(L_T, B, W) = \lceil k_{L} \cdot f(L_T) + k_{B} \cdot f(B) + k_{W} \cdot f(W) \rceil
\label{eq:line-f-def-affine}
\end{equation}

\begin{itemize}
	\item $k_{L} + k_{B} + k_{W} < 1 \implies $ the line is within the triangle
\end{itemize}

\noindent
Affine combination with context is in \xe{eq:line-f-def-affine-context}:

\begin{equation}
f(L_T, B, W) = \lceil k_{L} \cdot f(L_T) + k_{B} \cdot f(B) + k_{W} \cdot f(W) \rceil \pm \Delta c
\label{eq:line-f-def-affine-context}
\end{equation}

\noindent
where $\pm \Delta c$ is the amount of context surrounding the line, like in \tool{diff} \cite{diff};
with $c=0$ we are back to the original affine combination.

\paragraph{Learning approach with matrices and probabilities from examples.}

This case of the line number determination must follow the preliminary
positive test with some certainty that a give source code file contains
weaknesses and vulnerabilities. This methodology in itself would be
next to useless if this preliminary step is not performed.

In a simple case a line number is a cell in the 3D matrix $M$
given the file dimensions alone, as in \xe{eq:line-f-matrix-basic}. The matrix is sparse and
unknown entries are 0 by default. Non-zero entries are learned
from the examples of files with weaknesses. This matrix is
capable of encoding a single line location per file of the
same dimensions. As such it can't handle multiple locations
per file or two or more distinct unrelated files with different
line numbers for a single location. However, it serves as
a starting point to develop a further and better model.

\begin{equation}
l = f(L_T, B, W) = M[L_T, B, W]
\label{eq:line-f-matrix-basic}
\end{equation}

To allow multiple locations per file we either replace
the $W$ dimension with the locations dimension $N$ if
$W$ is not needed, as e.g. in \xe{eq:line-f-matrix-basic-vec},
or make the matrix 4D by adding $N$ to it, as in \xe{eq:line-f-matrix-basic-vec-n}.
This will take care of the multiple locations issue mentioned
earlier. $N$ is not known at the classification stage,
but the coordinates $L_T, B, W$ will give a value in the
3D matrix, which is a vector of locations $\vec{n}$.
At the reporting stage we simply report all of the elements
in $\vec{n}$.

\begin{equation}
\vec{l} = f(L_T, B, W) = M[L_T, B, N]
\label{eq:line-f-matrix-basic-vec}
\end{equation}

\begin{equation}
\vec{l} = f(L_T, B, W) = M[L_T, B, W, N]
\label{eq:line-f-matrix-basic-vec-n}
\end{equation}

In the above matrices $M$, the returned values are either a line number $l$
or a collection of line numbers $\vec{l}$ that were learned from examples
for the files of those dimensions. However, if we discovered a file tested
positive to contain a weakness, but we have never seen its dimensions (even
taking into the account we can sometimes ignore $W$), we'll get a zero.
This zero presents a problem: we can either (a) rely on one of the math functions
described earlier to fill in that zero with a non-zero line number or
(b) use probability values, and convert $M$ to $M_p$, as shown in \xe{eq:line-f-matrix-vec-p-n}.

The $M_p$ matrix would contain a vector value $\vec{n_p}$ of probabilities
a given line number is a line number of a weakness.

\begin{equation}
\vec{l_p} = f(L_T, B, W) = M_{p}[L_T, B, W, N]
\label{eq:line-f-matrix-vec-p-n}
\end{equation}

We then select the most probable ones from the list with the highest
probabilities. The index $i$ within $\vec{l_p}$ represents the line number and
the value at that index is the probability $p = \vec{l_p}[i]$.

Needless to say this 4D matrix is quite sparse and takes a while to learn.
The learning is performed by counting occurrences of line numbers of weaknesses
in the training data over total of entries.
To be better usable for the unseen cases the matrix needs to be smoothed using
any of the statistical estimators available, e.g. from NLP, such as add-delta,
ELE, MLE, Good-Turing, etc. by spreading the probabilities over to the zero-value
cells from the non-zero ones.
This is promising to be the slowest but the most accurate method.

In {\marf}, $M$ is implemented using \api{marf.util.Matrix},
a free-form matrix that grows upon the need lazily and allows
querying beyond physical dimensions when needed.

\subsubsection{Classes of Functions}

Define is the meaning of:

\begin{itemize}
	\item $k_?=\frac{L_T}{B}$
	\item $k_?=\frac{W}{B}$
\end{itemize}

Non-learning:

\begin{enumerate}
	\item 
		\begin{itemize}
			\item $k_*=1$
			\item $f(L_T) = L_T / 2$
			\item $f(B) = 0$
			\item $f(W) = 0$
		\end{itemize}
	\item 
		\begin{itemize}
			\item $k_L=\frac{W}{B}$
			\item $f(L_T) = L_T / 2$
			\item $f(B) = 0$
			\item $f(W) = 0$
		\end{itemize}
	\item 
		\begin{itemize}
			\item $k_L=\frac{L_T}{B}$
			\item $f(L_T) = L_T / 2$
			\item $f(B) = 0$
			\item $f(W) = 0$
		\end{itemize}
	\item 
		\begin{itemize}
			\item $k_*=1$
			\item $f(L_T) = random(L_T)$
			\item $f(B) = 0$
			\item $f(W) = 0$
		\end{itemize}
\end{enumerate}

\section{Results}
\label{sect:results}

The preliminary results of application of our methodology are outlined
in this section. We summarize the top precisions per test case using
either signal-processing or NLP-processing of the CVE-based cases and their
application to the general cases. Subsequent sections detail some of the
findings and issues of {\marfcat}'s result releases with different versions.

The results currently are being gradually released in the iterative manner
that were obtained through the corresponding versions of {\marfcat}
as it was being designed and developed.

\subsection{Preliminary Results Summary}
\label{sect:prelim-results-summary}

Current top precision at the SATE2010 timeframe:

\begin{itemize}
\item Wireshark:
	\begin{itemize}
	\item
	CVEs (signal): 92.68\%,
	CWEs (signal): 86.11\%,
	\item
	CVEs (NLP): 83.33\%,
	CWEs (NLP): 58.33\%
	\end{itemize}

\item Tomcat:
	\begin{itemize}
	\item
	CVEs (signal): 83.72\%,
	CWEs (signal): 81.82\%,
	\item
	CVEs (NLP): 87.88\%,
	CWEs (NLP): 39.39\%
	\end{itemize}
	
\item Chrome:
	\begin{itemize}
	\item
	CVEs (signal): 90.91\%,
	CWEs (signal): 100.00\%,
	\item
	CVEs (NLP): 100.00\%,
	CWEs (NLP): 88.89\%
	\end{itemize}

\item Dovecot:
	\begin{itemize}
	\item
	14 warnings; but it appears all quality or false positive
	\item
	(very hard to follow the code, severely undocumented)
	\end{itemize}

\item Pebble:
	\begin{itemize}
	\item
	none found during quick testing
	\end{itemize}

\end{itemize}

What follows are some select statistical measurements of the
precision in recognizing CVEs and CWEs under different configurations
using the signal processing and NLP processing techniques.

``Second guess'' statistics provided to see if the hypothesis that
if our first estimate of a CVE/CWE is incorrect, the next one in line
is probably the correct one. Both are counted if the first guess is
correct.

\subsection{Version SATE.4}

\subsubsection{\wiresharkCase{0}}

Typical quick run on the enriched \wiresharkCase{0} on CVEs is in \xt{tab:wireshark120-sate4-cves-quick-enriched}.
All 22 CVEs are reported. Pretty good precision for options \option{-diff}
and \option{-cheb} (Diff and Chebyshev distance classifiers, respectively \cite{marf-c3s2e08}).
In Unigram, Add-Delta NLP results on
\wiresharkCase{0}'s
training file for CVEs, the
precision seems to be overall degraded compared
to the classical signal processing pipeline. Only 20 out of 22 CVEs are reported,
as shown in \xt{tab:wireshark120-sate4-nlp-cves-quick-enriched}.
CWE-based testing on
\wiresharkCase{0} (also with
some basic line heuristics that does not impact the
precision) is in \xt{tab:wireshark120-sate4-cwes-quick-enriched}.

The following select reports are about \wiresharkCase{0} using a small subset
of algorithms.
There are line numbers that were machine-learned from the \texttt{\_train.xml} file.
The two
XML report
files are the best ones we have chosen
among several of them. Their precision rate using machine learning techniques
is 92.68\% after several bug corrections done.
All CVEs are reported
making recall 100\%.
The \texttt{stats-*.txt} files are there
summarizing the evaluation precision. The results are as good as
the training data given; if there are mistakes in the data selection
and annotation XML files, then the results will also have mistakes
accordingly.

\noindent
The best reports are:

\noindent
  \file{report-noprepreprawfftcheb-wireshark-1.2.0-train.xml}
	
\noindent
  \file{report-noprepreprawfftdiff-wireshark-1.2.0-train.xml}

\noindent
The first one validates with both sate2010 schemas, but
the latter has problems with the exponential -E notation.

\paragraph*{Files.}

The corresponding \texttt{*.log} files are there for references,
but contain a lot of debug information from the tool.
The tool is using thresholding to reduce the amount of noise
going into the reports.

\begin{verbatim}
marfcat-nopreprep-raw-fft-cheb.log
marfcat-nopreprep-raw-fft-diff.log
marfcat--super-fast.log (primaily training log)
report-noprepreprawfftcheb-wireshark-1.2.0-train.xml
report-noprepreprawfftdiff-wireshark-1.2.0-train.xml
stats--super-fast.txt
wireshark-1.2.0_train.xml
\end{verbatim}

\subsubsection{\wiresharkCase{9}}

The following analysis reports are about \wiresharkCase{9} using a small subset
of {\marf}'s algorithms.
The system correctly does {\em not} report the fixed CVEs (currently,
the primary class), so most of the reports come up empty (no noise).
All example reports (one per configuration) validate with the schemas
\file{sate\_2010.xsd} and \file{sate\_2010.pathcheck.xsd}.

\noindent
The best (empty) reports are:

\noindent
  \file{report-noprepreprawfftcheb-wireshark-1.2.9-test.xml}
	
\noindent
  \file{report-noprepreprawfftdiff-wireshark-1.2.9-test.xml}

\noindent
  \file{report-noprepreprawffteucl-wireshark-1.2.9-test.xml}
	
\noindent
  \file{report-noprepreprawffthamming-wireshark-1.2.9-test.xml}

The below particular report shows the Minkowski distance classifier (\option{-mink})
was not perhaps the best choice, as it mistakingly reported
a known CVE that was in fact fixed, this is an example of
machine learning ``red herring'':

\noindent
  \file{report-noprepreprawfftmink-wireshark-1.2.9-test.xml}

\paragraph*{Files.}

All the corresponding tool-specific *.log files are there for reference.

\begin{verbatim}
marfcat-nopreprep-raw-fft-cheb.log
marfcat-nopreprep-raw-fft-diff.log
marfcat-nopreprep-raw-fft-eucl.log
marfcat-nopreprep-raw-fft-hamming.log
marfcat-nopreprep-raw-fft-mink.log
marfcat--super-fast-wireshark.log (training log)
report-noprepreprawfftcheb-wireshark-1.2.9-test.xml
report-noprepreprawfftdiff-wireshark-1.2.9-test.xml
report-noprepreprawffteucl-wireshark-1.2.9-test.xml
report-noprepreprawffthamming-wireshark-1.2.9-test.xml
report-noprepreprawfftmink-wireshark-1.2.9-test.xml
\end{verbatim}

\subsubsection{\chromeCase{54}}

This version's CVE testing result of \chromeCase{54} (after updates and removal unrelated
CVEs per SATE organizers) is in \xt{tab:chrome54-sate4-cves-quick-enriched-clean-cvs}.
The corresponding select reports produced below are
about \chromeCase{54} using a small subset of algorithms.
There are line numbers that were machine-learned from the \texttt{*\_train.xml} file.
The two
\texttt{report-*.xml} files are ones of the best ones we have picked. 
Their precision rate using machine learning techniques is 90.91\% after all
the corrections done. The \texttt{stats-*.txt} file is there summarizing the evaluation
precision in the end of that file. Again, the results are as good as
the training data given; if there are mistakes in the data selection
and annotation XML files, then the results will also have mistakes
accordingly.

\noindent
The best reports are:

\noindent
  \file{report-noprepreprawfftcheb-chrome-5.0.375.54-train.xml}
	
\noindent
  \file{report-noprepreprawfftdiff-chrome-5.0.375.54-train.xml}

\noindent
Both validate with both sate2010 schemas.

\paragraph*{Files.}

The corresponding \texttt{*.log} files are there for references, but contain A LOT
of debug info from the tool. The tool is using thresholding to
reduce the amount of noise going into the reports, but if you
are curious to examine the logs, they are included.

\begin{verbatim}
chrome-5.0.375.54_train.xml
marfcat-nopreprep-raw-fft-cheb.log
marfcat-nopreprep-raw-fft-diff.log
marfcat--super-fast-chrome.log
README.txt
report-noprepreprawfftcheb-chrome-5.0.375.54-train.xml
report-noprepreprawfftdiff-chrome-5.0.375.54-train.xml
stats--super-fast.txt
\end{verbatim}

\subsubsection{\chromeCase{70}}

The following reports are about \chromeCase{70} using a small subset
of algorithms.
The system correctly does {\em not} report the fixed CVEs, 
so most of the reports come up empty (no noise) as they are
expected to be for known CVE-selected weaknesses.
All example reports (one per configuration) validate with the schema
\file{sate\_2010.xsd} and \file{sate\_2010.pathcheck.xsd}.

\noindent
The best (empty) reports are:

\noindent
  \file{report-noprepreprawfftcheb-chrome-5.0.375.70-test.xml}
	
\noindent
  \file{report-noprepreprawfftdiff-chrome-5.0.375.70-test.xml}
	
\noindent
  \file{report-noprepreprawffteucl-chrome-5.0.375.70-test.xml}
	
\noindent
  \file{report-noprepreprawffthamming-chrome-5.0.375.70-test.xml}
	
\noindent
  \file{report-noprepreprawfftmink-chrome-5.0.375.70-test.xml}

\paragraph*{Files.}

All the corresponding tool-specific \texttt{*.log} files are there for reference.

\begin{verbatim}
chrome-5.0.375.70_test.xml
marfcat-nopreprep-raw-fft-cheb.log
marfcat-nopreprep-raw-fft-diff.log
marfcat-nopreprep-raw-fft-eucl.log
marfcat-nopreprep-raw-fft-hamming.log
marfcat-nopreprep-raw-fft-mink.log
marfcat--super-fast-chrome.log
report-noprepreprawfftcheb-chrome-5.0.375.70-test.xml
report-noprepreprawfftdiff-chrome-5.0.375.70-test.xml
report-noprepreprawffteucl-chrome-5.0.375.70-test.xml
report-noprepreprawffthamming-chrome-5.0.375.70-test.xml
report-noprepreprawfftmink-chrome-5.0.375.70-test.xml
\end{verbatim}

\subsection{Version SATE.5}

\subsubsection{\chromeCase{54}}

Here we complete the CVE results from the {\marfcat} SATE.5 version by
using \chromeCase{54} training on \chromeCase{54} with
classical CWEs as opposed to CVEs. The result summary 
is in \xt{tab:chrome54-sate5-cwes}.

\subsubsection{\tomcatCase{13}}

With this {\marfcat} version we did
first
CVE-based testing on training for \tomcatCase{13}. 
Classifiers corresponding to \option{-cheb} (Chebyshev distance) and \option{-diff} (Diff distance)
continue to dominate as in the other test cases.
An observation: for some reason, \option{-cos} (cosine similarity
classifier) with the same settings as for the {\C}/{\cpp} projects (Wireshark and Chrome)
actually preforms well and \texttt{*\_report.xml} is not as noisy; in fact
comparable to \option{-cheb} and \option{-diff}.
These CVE-based results are summarized in \xt{tab:tomcat13-sate5-cves}.
Further, we perform quick CWE-based testing on \tomcatCase{13}. Reports are quite
larger for \option{-cheb}, \option{-diff}, and \option{-cos},
but not for other classifiers. The precision results are
illustrated in \xt{tab:tomcat13-sate5-cwes}.
Then, in SATE.5, quick \tomcatCase{13} CVE NLP testing shows higher precision of 87.88\%,
but the recall is poor, 25/31 -- 6 CVEs are missing out (see \xt{tab:tomcat13-sate5-nlp-cves}).
Subsequent, quick \tomcatCase{13} CWE NLP testing was surprisingly poor topping
at 39.39\% (see \xt{tab:tomcat13-sate5-nlp-cwes}).
The resulting select reports about this Apache \tomcatCase{13} test case
using a small subset of algorithms are mentioned below with some commentary.

\paragraph*{CVE-based training and reporting:}

As before, there are line numbers that were machine-learned from the \texttt{\_train.xml} file
as well as the types of locations and descriptions provided by the SATE
organizers and incorporated into the reports via machine learning.
This includes the types of locations, such as ``fix'', ``sink'', or ``path''
learned from the ogranizers-provided XML/spreadsheet as well as the source code files.
Two of all the produced
XML reports are the best ones. The macro precision
rate in there using machine learning techniques is 83.72\%. The \texttt{stats-*.txt} files are there
summarizing the evaluation precision.

\noindent
The best reports are:

\noindent
  \file{report-noprepreprawfftcheb-apache-tomcat-5.5.13-train-cve.xml}
	
\noindent
  \file{report-noprepreprawfftdiff-apache-tomcat-5.5.13-train-cve.xml}\\(does not validate three tool-specific lines)

\noindent
Other reports are, to a various degree of detail and noise:

\noindent
  \file{report-noprepreprawfftcos-apache-tomcat-5.5.13-train-cve.xml}\\(does not validate two lines)
	
\noindent
  \file{report-noprepreprawffteucl-apache-tomcat-5.5.13-train-cve.xml}\\(does not validate three tool-specific lines)
	
\noindent
  \file{report-noprepreprawffthamming-apache-tomcat-5.5.13-train-cve.xml}
	
\noindent
  \file{report-noprepreprawfftmink-apache-tomcat-5.5.13-train-cve.xml}
	
\noindent
  \file{report-nopreprepcharunigramadddelta-apache-tomcat-5.5.13-train-cve-nlp.xml}

The \option{--nlp} version reports use the NLP techniques with the machine
learning instead of signal processing techniques. Those reports
are largely comparable, but have smaller recall, 
i.e. some CVEs are completely missing out from the reports in this version. 
Some reports have problems with tool-specific ranks like:
$4.199735736674989E-4$, which we will have to see how to reduce these.

\paragraph*{CWE-based training and reporting:}

The CWE-based reports use the CWE as a primary class instead of CVE
for training and reporting, and as such currently do not report on
CVEs directly (i.e. no direct mapping from CWE to CVE exists unlike
in the opposite direction); however, their recognition rates are not very low
either in the same spots, types, etc. In the future version of
{\marfcat} the plan is to combine the two machine learning pipeline
runs of CVE and CWE together to improve mutual classification,
but right now it is not available. The CWE-based training is also
used on the testing files say of {\pebbleCase} to see if there are any
similar weaknesses to that of Tomcat found, again e.g. in Pebble. CWEs,
unlike CVEs for most projects, represent better cross-project
classes as they are largely project-independent. Both CVE-based
and CWE-base methods use the same data for training. CWEs are
recognized correctly 81.82\% for Tomcat. NLP-based CWE testing
is not included as its precision was quite low ($\approx 39\%$).

\noindent
The best reports are:

\noindent
  \file{report-cweidnoprepreprawfftcheb-apache-tomcat-5.5.13-train-cwe.xml}\\(does not validate)
	
\noindent
  \file{report-cweidnoprepreprawfftdiff-apache-tomcat-5.5.13-train-cwe.xml}\\(does not validate)

\noindent
Other reports are, to a various degree of detail and noise:

\noindent
  \file{report-cweidnoprepreprawfftcos-apache-tomcat-5.5.13-train-cwe.xml}
	
\noindent
  \file{report-cweidnoprepreprawffteucl-apache-tomcat-5.5.13-train-cwe.xml}\\(does not validate)
	
\noindent
  \file{report-cweidnoprepreprawffthamming-apache-tomcat-5.5.13-train-cwe.xml}
	
\noindent
  \file{report-cweidnoprepreprawfftmink-apache-tomcat-5.5.13-train-cwe.xml}

\paragraph*{Files.}

The corresponding *.log files are there for references, but contain A LOT
of debug info from the tool. The tool is using thresholding to
reduce the amount of noise going into the reports, but if you
are curious to examine the logs, they are included.

\begin{verbatim}
apache-tomcat-5.5.13-src_train.xml (meta training file)
marfcat-cweid-nopreprep-raw-fft-cheb.log
marfcat-cweid-nopreprep-raw-fft-cos.log
marfcat-cweid-nopreprep-raw-fft-diff.log
marfcat-cweid-nopreprep-raw-fft-eucl.log
marfcat-cweid-nopreprep-raw-fft-hamming.log
marfcat-cweid-nopreprep-raw-fft-mink.log
marfcat-nopreprep-char-unigram-add-delta.log
marfcat-nopreprep-raw-fft-cheb.log
marfcat-nopreprep-raw-fft-cos.log
marfcat-nopreprep-raw-fft-diff.log
marfcat-nopreprep-raw-fft-eucl.log
marfcat-nopreprep-raw-fft-hamming.log
marfcat-nopreprep-raw-fft-mink.log
marfcat--super-fast-tomcat-train-cve.log
marfcat--super-fast-tomcat-train-cve-nlp.log
marfcat--super-fast-tomcat-train-cwe.log
report-cweidnoprepreprawfftcheb-apache-tomcat-5.5.13-train-cwe.xml
report-cweidnoprepreprawfftcos-apache-tomcat-5.5.13-train-cwe.xml
report-cweidnoprepreprawfftdiff-apache-tomcat-5.5.13-train-cwe.xml
report-cweidnoprepreprawffteucl-apache-tomcat-5.5.13-train-cwe.xml
report-cweidnoprepreprawffthamming-apache-tomcat-5.5.13-train-cwe.xml
report-cweidnoprepreprawfftmink-apache-tomcat-5.5.13-train-cwe.xml
report-nopreprepcharunigramadddelta-apache-tomcat-5.5.13-train-cve-nlp.xml
report-noprepreprawfftcheb-apache-tomcat-5.5.13-train-cve.xml
report-noprepreprawfftcos-apache-tomcat-5.5.13-train-cve.xml
report-noprepreprawfftdiff-apache-tomcat-5.5.13-train-cve.xml
report-noprepreprawffteucl-apache-tomcat-5.5.13-train-cve.xml
report-noprepreprawffthamming-apache-tomcat-5.5.13-train-cve.xml
report-noprepreprawfftmink-apache-tomcat-5.5.13-train-cve.xml
stats-per-cve-nlp.txt
stats-per-cve.txt
stats-per-cwe.txt
\end{verbatim}

\subsubsection{{\pebbleCase} 2.5-M2}

Using the machine learning approach of {\marf} by using the \tomcatCase{13} as
a source of training on a {\java} project with known weaknesses, we used
that (rather small) ``knowledge base'' to test if anything weak similar
to the weaknesses in Tomcat are also present in the supplied version
of {\pebbleCase} 2.5-M2. The current result is that under the version of {\marfcat} SATE.5
all reports come up empty under the current thresholding rules meaning
the tool was not able to identify similar weaknesses in files in Pebble.
The corresponding tool-specific log files are also provided if of interest,
but the volume of data in them is typically large.
It is planned
to lower the thresholds
after reviewing logs in detail to see if anything interesting comes up
that we missed otherwise.

\paragraph*{Files.}

\begin{verbatim}
marfcat--super-fast-tomcat13-pebble-cwe.log
marfcat-cweid-nopreprep-raw-fft-cheb.log
marfcat-cweid-nopreprep-raw-fft-cos.log
marfcat-cweid-nopreprep-raw-fft-diff.log
marfcat-cweid-nopreprep-raw-fft-eucl.log
marfcat-cweid-nopreprep-raw-fft-hamming.log
marfcat-cweid-nopreprep-raw-fft-mink.log
report-cweidnoprepreprawfftcheb-pebble-test-cwe.xml
report-cweidnoprepreprawfftcos-pebble-test-cwe.xml
report-cweidnoprepreprawfftdiff-pebble-test-cwe.xml
report-cweidnoprepreprawffteucl-pebble-test-cwe.xml
report-cweidnoprepreprawffthamming-pebble-test-cwe.xml
report-cweidnoprepreprawfftmink-pebble-test-cwe.xml
\end{verbatim}

\subsubsection{Tomcat and Pebble Testing Results Summary}

\begin{itemize}
	\item 
\tomcatCase{13} on \tomcatCase{29} classical CVE testing produced only report
with \option{-cos} with 10 weaknesses, some correspond to the files in
training. However, the line numbers reported are midline, so
next to meaningless.

	\item 
\tomcatCase{13} on \tomcatCase{29} classical CWE testing also report with \option{-cos}
with 2 weaknesses.

	\item 
\tomcatCase{13} on \tomcatCase{29} NLP CVE testing single report (quick testing
only does add-delta, unigram) came up empty.

	\item 
\tomcatCase{13} on \tomcatCase{29} NLP CWE testing, also with a single report (quick testing
only does add-delta, unigram) came up empty.

	\item 
\tomcatCase{13} on {\pebbleCase} classical CVE reports are empty.

	\item 
\tomcatCase{13} on Pebble NLP CVE report is not empty, but reports
wrongly on \texttt{blank.html} (empty HTML file) on multiple CVEs.
The probability $P=0.0$ for all in this case CVEs, not sure why it is at all reported.
A red herring.

	\item 
\tomcatCase{13} on {\pebbleCase} classical CWE reports are empty.

	\item 
\tomcatCase{13} on {\pebbleCase} NLP CWE is similar to the Pebble NLP CVE
report on \texttt{blank.html} entries, but fewer of them. All the
other symptoms are the same.
\end{itemize}

\subsection{Version SATE.6}

\subsubsection{{\dovecotCase} 2.0.beta6}

This is a quick
test and a report
for {\dovecotCase} 2.0.beta6, with line numbers and other information.
The report is `raw', without our manual evaluation and generated as-is
at this point.

\noindent
The report of interest:

\noindent
  \file{report-cweidnoprepreprawfftcos-dovecot-2.0.beta6-wireshark-test-cwe.xml}

\noindent
It appears though from the first glance most of the are warnings are `bogus' or
`buggy', but could indicate potential presence of weaknesses in the
flagged files. One thing is for sure the Dovecode's source code's main
weakness is a near chronic lack of comments, which is also a weakness of
a kind.
Other reports came up empty. The source for learning was \wiresharkCase{0}.

\paragraph*{Files.}

\begin{verbatim}
dovecot-2.0.beta6_test.xml
marfcat--super-fast-dovecot-wireshark-test-cwe.log
marfcat-cweid-nopreprep-raw-fft-cheb.log
marfcat-cweid-nopreprep-raw-fft-cos.log
marfcat-cweid-nopreprep-raw-fft-diff.log
marfcat-cweid-nopreprep-raw-fft-eucl.log
marfcat-cweid-nopreprep-raw-fft-hamming.log
marfcat-cweid-nopreprep-raw-fft-mink.log
report-cweidnoprepreprawfftcheb-dovecot-2.0.beta6-wireshark-test-cwe.xml
report-cweidnoprepreprawfftcheb-wireshark-1.2.0_train.xml.xml
report-cweidnoprepreprawfftcos-dovecot-2.0.beta6-wireshark-test-cwe.xml
report-cweidnoprepreprawfftdiff-dovecot-2.0.beta6-wireshark-test-cwe.xml
report-cweidnoprepreprawffteucl-dovecot-2.0.beta6-wireshark-test-cwe.xml
report-cweidnoprepreprawffthamming-dovecot-2.0.beta6-wireshark-test-cwe.xml
report-cweidnoprepreprawfftmink-dovecot-2.0.beta6-wireshark-test-cwe.xml
\end{verbatim}

\subsubsection{\tomcatCase{29}}

This is another quick
CVE-based evaluation
of \tomcatCase{29}, with line numbers, etc.
They are 'raw', without our manual evaluation and generated as-is.

\noindent
The reports of interest:

\noindent
  \file{report-noprepreprawfftcos-apache-tomcat-5.5.29-test-cve.xml}
	
\noindent
  \file{report-cweidnoprepreprawfftcos-apache-tomcat-5.5.29-test-cwe.xml}

\noindent
As for the Dovecot case,
it appears though from the first glance most of the warnings are either `bogus' or
`buggy', but could indicate potential presence of weaknesses in the
flagged files or fixed as such. Need more manual inspection to be sure.
Other XML reports came up empty. The source for learning was \tomcatCase{13}.

\paragraph*{Files.}

\begin{verbatim}
marfcat--super-fast-tomcat13-tomcat29-cve.log
marfcat--super-fast-tomcat13-tomcat29-cwe.log
marfcat-cweid-nopreprep-raw-fft-cheb.log
marfcat-cweid-nopreprep-raw-fft-cos.log
marfcat-cweid-nopreprep-raw-fft-diff.log
marfcat-cweid-nopreprep-raw-fft-eucl.log
marfcat-cweid-nopreprep-raw-fft-hamming.log
marfcat-cweid-nopreprep-raw-fft-mink.log
marfcat-nopreprep-raw-fft-cheb.log
marfcat-nopreprep-raw-fft-cos.log
marfcat-nopreprep-raw-fft-diff.log
marfcat-nopreprep-raw-fft-eucl.log
marfcat-nopreprep-raw-fft-hamming.log
marfcat-nopreprep-raw-fft-mink.log
report-cweidnoprepreprawfftcheb-apache-tomcat-5.5.29-test-cwe.xml
report-cweidnoprepreprawfftcos-apache-tomcat-5.5.29-test-cwe.xml
report-cweidnoprepreprawfftdiff-apache-tomcat-5.5.29-test-cwe.xml
report-cweidnoprepreprawffteucl-apache-tomcat-5.5.29-test-cwe.xml
report-cweidnoprepreprawffthamming-apache-tomcat-5.5.29-test-cwe.xml
report-cweidnoprepreprawfftmink-apache-tomcat-5.5.29-test-cwe.xml
report-noprepreprawfftcheb-apache-tomcat-5.5.29-test-cve.xml
report-noprepreprawfftcos-apache-tomcat-5.5.29-test-cve.xml
report-noprepreprawfftdiff-apache-tomcat-5.5.29-test-cve.xml
report-noprepreprawffteucl-apache-tomcat-5.5.29-test-cve.xml
report-noprepreprawffthamming-apache-tomcat-5.5.29-test-cve.xml
report-noprepreprawfftmink-apache-tomcat-5.5.29-test-cve.xml
\end{verbatim}

\subsection{Version SATE.7}

Up until this version NLP processing of Chrome was not successful.
Errors related to the number of file descriptors opened and ``mark
invalid'' for NLP processing of \chromeCase{54} for both CVEs and CWEs
have been corrected, so we have produced the results for these cases.
CVEs are reported in \xt{tab:chrome54-sate7-nlp-cves}.
CWEs are further reported in \xt{tab:chrome54-sate7-nlp-cwes}.

\section{Conclusion}
\label{sect:conclusion}

We review the current results of this experimental work, its current shortcomings,
advantages, and practical implications. We also release {\marfcat} Alpha version
as open-source that can be found at \cite{marfcat-app}. This is following the
open-source philosophy of greater good ({\marf} itself has been open-source
from the very beginning \cite{marf}).

\subsection{Shortcomings}
\label{sect:shortcomings}

The below is a list of most prominent issues with the presented approach.
Some of them are more ``permanent'', while others are solvable and intended to
be addressed in the future work. Specifically:

\begin{itemize}
\item
Looking at a signal is less intuitive visually for code analysis by humans.

\item
Line numbers are a problem (easily ``filtered out'' as high-frequency ``noise'', etc.). A whole ``relativistic''
and machine learning methodology developed for the line numbers in \xs{sect:line-numbers} to compensate for that.
Generally, when CVEs is the primary class, by accurately identifying the CVE number one can get all the
other pertinent details from the CVE database, including patches and line numbers.

\item
Accuracy depends on the quality of the knowledge base (see \xs{sect:kb}) collected.
``Garbage in -- garbage out.''

\item
To detect CVE or CWE signatures in non-CVE cases requires large knowledge
bases (human-intensive to collect).

\item
No path tracing (since no parsing is present); no slicing, semantic annotations,
context, locality of reference, etc. The ``sink'', ``path'', and ``fix'' results
in the reports also have to be machine-learned.

\item
A lot of algorithms and their combinations to try (currently $\approx 1800$ permutations)
to get the best top N. This is, however, also an advantage of the approach as the
underlying framework can quickly allow for such testing.

\item
File-level training vs. fragment-level training -- presently the classes are trained
based on the entire file where weaknesses are found instead of the known fragments from
CVE-reported patches. The latter would be more fine-grained and precise than whole-file
classification, but slower. However, overall the file-level processing is a man-hour
limitation than a technological one.

\item
No nice GUI. Presently the application is script/command-line based.
\end{itemize}

\subsection{Advantages}

There are some key advantages of the approach presented. Some of them
follow:

\begin{itemize}
\item
Relatively fast (e.g. Wireshark's $\approx 2400$ files train and test
in about 3 minutes) on a now-commodity desktop.

\item
Language-independent (no parsing) -- given enough examples can apply to any language,
i.e. methodology is the same no matter {\C}, {\cpp}, {\java} or any other source
or binary languages (PHP, C\#, VB, Perl, bytecode, assembly, etc.).

\item
Can automatically learn a large knowledge base to test on known and unknown cases.

\item
Can be used to quickly pre-scan projects for further analysis by humans
and other tools that do in-depth semantic analysis.

\item
Can learn from other SATE'10 reports.

\item
Can learn from SATE'09 and SATE'08 reports.

\item
High precision in CVEs and CWE detection.

\item
Lots of algorithms and their combinations to select the best
for a particular task or class (see \xs{sect:ml-cats}).
\end{itemize}

\subsection{Practical Implications}

Most practical implications of all static code analyzers are obvious -- to
detect and report source code weaknesses and report them appropriately to
the developers. We outline additional implications this approach brings
to the arsenal below:

\begin{itemize}

\item
The approach can be used on any target language without modifications to the
methodology or knowing the syntax of the language. Thus, it scales to any
popular and new language analysis with a very small amount of effort.

\item
The approach can nearly identically be transposed onto the compiled binaries and bytecode, detecting
vulnerable deployments and installations -- sort of like virus scanning of
binaries, but instead scanning for infected binaries, one would scan for security-weak binaries
on site deployments to alert system administrators to upgrade their packages.

\item
Can learn from binary signatures from other tools like Snort \cite{snort}.
\end{itemize}

\subsection{Future Work}

There is a great number of possibilities in the future work. This includes
improvements to the code base of {\marfcat} as well as resolving unfinished
scenarios and results, addressing shortcomings in \xs{sect:shortcomings},
testing more algorithms and combinations from the related work,
and moving onto other programming languages (e.g. PHP, ASP, C\#).
Furthermore, plan to conceive collaboration with vendors such as VeraCode, Coverity,
and others who have vast data sets to test the full potential of the approach
with the others and a community as a whole. Then move on to dynamic code
analysis as well applying similar techniques there.

\bibliographystyle{alpha}
\NIST{}
\bibliography{marf-sate}

\appendix

\section{Classification Result Tables}
\label{sect:classification-results}
\label{appdx:classification-results}

What follows are result tables with top classification results
ranked from most precise at the top. This include the configuration
settings for {\marf} by the means of options (the algorithm implementations
are at their defaults \cite{marf-cisse07}).

\begin{table}%
\centering
\caption{CVE Stats for \wiresharkCase{0}, Quick Enriched, version SATE.4}
\label{tab:wireshark120-sate4-cves-quick-enriched}
\scriptsize
\begin{tabular}{|c|r|l|c|c|r|}\hline
guess & run & algorithms & good & bad &  \% \\\hline
1st & 1 & \option{-nopreprep} \option{-raw} \option{-fft} \option{-diff}  & 38 & 3 & 92.68\\\hline
1st & 2 & \option{-nopreprep} \option{-raw} \option{-fft} \option{-cheb}  & 38 & 3 & 92.68\\\hline
1st & 3 & \option{-nopreprep} \option{-raw} \option{-fft} \option{-eucl}  & 29 & 12 & 70.73\\\hline
1st & 4 & \option{-nopreprep} \option{-raw} \option{-fft} \option{-hamming}  & 26 & 15 & 63.41\\\hline
1st & 5 & \option{-nopreprep} \option{-raw} \option{-fft} \option{-mink}  & 23 & 18 & 56.10\\\hline
1st & 6 & \option{-nopreprep} \option{-raw} \option{-fft} \option{-cos}  & 37 & 51 & 42.05\\\hline
2nd & 1 & \option{-nopreprep} \option{-raw} \option{-fft} \option{-diff}  & 39 & 2 & 95.12\\\hline
2nd & 2 & \option{-nopreprep} \option{-raw} \option{-fft} \option{-cheb}  & 39 & 2 & 95.12\\\hline
2nd & 3 & \option{-nopreprep} \option{-raw} \option{-fft} \option{-eucl}  & 34 & 7 & 82.93\\\hline
2nd & 4 & \option{-nopreprep} \option{-raw} \option{-fft} \option{-hamming}  & 28 & 13 & 68.29\\\hline
2nd & 5 & \option{-nopreprep} \option{-raw} \option{-fft} \option{-mink}  & 31 & 10 & 75.61\\\hline
2nd & 6 & \option{-nopreprep} \option{-raw} \option{-fft} \option{-cos}  & 38 & 50 & 43.18\\\hline
guess & run & class & good & bad &  \% \\\hline
1st & 1 & \cve{CVE-2009-3829} & 6 & 0 & 100.00\\\hline
1st & 2 & \cve{CVE-2009-2563} & 6 & 0 & 100.00\\\hline
1st & 3 & \cve{CVE-2009-2562} & 6 & 0 & 100.00\\\hline
1st & 4 & \cve{CVE-2009-4378} & 6 & 0 & 100.00\\\hline
1st & 5 & \cve{CVE-2009-4376} & 6 & 0 & 100.00\\\hline
1st & 6 & \cve{CVE-2010-0304} & 6 & 0 & 100.00\\\hline
1st & 7 & \cve{CVE-2010-2286} & 6 & 0 & 100.00\\\hline
1st & 8 & \cve{CVE-2010-2283} & 6 & 0 & 100.00\\\hline
1st & 9 & \cve{CVE-2009-3551} & 6 & 0 & 100.00\\\hline
1st & 10 & \cve{CVE-2009-3550} & 6 & 0 & 100.00\\\hline
1st & 11 & \cve{CVE-2009-3549} & 6 & 0 & 100.00\\\hline
1st & 12 & \cve{CVE-2009-3241} & 16 & 8 & 66.67\\\hline
1st & 13 & \cve{CVE-2010-1455} & 34 & 20 & 62.96\\\hline
1st & 14 & \cve{CVE-2009-3243} & 18 & 11 & 62.07\\\hline
1st & 15 & \cve{CVE-2009-2560} & 8 & 6 & 57.14\\\hline
1st & 16 & \cve{CVE-2009-2561} & 6 & 5 & 54.55\\\hline
1st & 17 & \cve{CVE-2010-2285} & 6 & 5 & 54.55\\\hline
1st & 18 & \cve{CVE-2009-2559} & 6 & 5 & 54.55\\\hline
1st & 19 & \cve{CVE-2010-2287} & 6 & 6 & 50.00\\\hline
1st & 20 & \cve{CVE-2009-4377} & 12 & 15 & 44.44\\\hline
1st & 21 & \cve{CVE-2010-2284} & 6 & 9 & 40.00\\\hline
1st & 22 & \cve{CVE-2009-3242} & 7 & 12 & 36.84\\\hline
2nd & 1 & \cve{CVE-2009-3829} & 6 & 0 & 100.00\\\hline
2nd & 2 & \cve{CVE-2009-2563} & 6 & 0 & 100.00\\\hline
2nd & 3 & \cve{CVE-2009-2562} & 6 & 0 & 100.00\\\hline
2nd & 4 & \cve{CVE-2009-4378} & 6 & 0 & 100.00\\\hline
2nd & 5 & \cve{CVE-2009-4376} & 6 & 0 & 100.00\\\hline
2nd & 6 & \cve{CVE-2010-0304} & 6 & 0 & 100.00\\\hline
2nd & 7 & \cve{CVE-2010-2286} & 6 & 0 & 100.00\\\hline
2nd & 8 & \cve{CVE-2010-2283} & 6 & 0 & 100.00\\\hline
2nd & 9 & \cve{CVE-2009-3551} & 6 & 0 & 100.00\\\hline
2nd & 10 & \cve{CVE-2009-3550} & 6 & 0 & 100.00\\\hline
2nd & 11 & \cve{CVE-2009-3549} & 6 & 0 & 100.00\\\hline
2nd & 12 & \cve{CVE-2009-3241} & 17 & 7 & 70.83\\\hline
2nd & 13 & \cve{CVE-2010-1455} & 44 & 10 & 81.48\\\hline
2nd & 14 & \cve{CVE-2009-3243} & 18 & 11 & 62.07\\\hline
2nd & 15 & \cve{CVE-2009-2560} & 9 & 5 & 64.29\\\hline
2nd & 16 & \cve{CVE-2009-2561} & 6 & 5 & 54.55\\\hline
2nd & 17 & \cve{CVE-2010-2285} & 6 & 5 & 54.55\\\hline
2nd & 18 & \cve{CVE-2009-2559} & 6 & 5 & 54.55\\\hline
2nd & 19 & \cve{CVE-2010-2287} & 12 & 0 & 100.00\\\hline
2nd & 20 & \cve{CVE-2009-4377} & 12 & 15 & 44.44\\\hline
2nd & 21 & \cve{CVE-2010-2284} & 6 & 9 & 40.00\\\hline
2nd & 22 & \cve{CVE-2009-3242} & 7 & 12 & 36.84\\\hline
\end{tabular}

\normalsize
\end{table}

\begin{table}%
\centering
\caption{CVE NLP Stats for \wiresharkCase{0}, Quick Enriched, version SATE.4}
\label{tab:wireshark120-sate4-nlp-cves-quick-enriched}
\begin{tabular}{|c|r|l|c|c|r|}\hline
guess & run & algorithms & good & bad &  \% \\\hline
1st & 1 & \option{-nopreprep} \option{-char} \option{-unigram} -add-delta  & 30 & 6 & 83.33\\\hline
2nd & 1 & \option{-nopreprep} \option{-char} \option{-unigram} -add-delta  & 31 & 5 & 86.11\\\hline
guess & run & class & good & bad &  \% \\\hline
1st & 1 & \cve{CVE-2009-3829} & 1 & 0 & 100.00\\\hline
1st & 2 & \cve{CVE-2009-2563} & 1 & 0 & 100.00\\\hline
1st & 3 & \cve{CVE-2009-2562} & 1 & 0 & 100.00\\\hline
1st & 4 & \cve{CVE-2009-4378} & 1 & 0 & 100.00\\\hline
1st & 5 & \cve{CVE-2009-2561} & 1 & 0 & 100.00\\\hline
1st & 6 & \cve{CVE-2009-4377} & 1 & 0 & 100.00\\\hline
1st & 7 & \cve{CVE-2009-4376} & 1 & 0 & 100.00\\\hline
1st & 8 & \cve{CVE-2010-2286} & 1 & 0 & 100.00\\\hline
1st & 9 & \cve{CVE-2010-0304} & 1 & 0 & 100.00\\\hline
1st & 10 & \cve{CVE-2010-2285} & 1 & 0 & 100.00\\\hline
1st & 11 & \cve{CVE-2010-2284} & 1 & 0 & 100.00\\\hline
1st & 12 & \cve{CVE-2010-2283} & 1 & 0 & 100.00\\\hline
1st & 13 & \cve{CVE-2009-2559} & 1 & 0 & 100.00\\\hline
1st & 14 & \cve{CVE-2009-3550} & 1 & 0 & 100.00\\\hline
1st & 15 & \cve{CVE-2009-3549} & 1 & 0 & 100.00\\\hline
1st & 16 & \cve{CVE-2010-1455} & 8 & 1 & 88.89\\\hline
1st & 17 & \cve{CVE-2009-3243} & 3 & 1 & 75.00\\\hline
1st & 18 & \cve{CVE-2009-3241} & 2 & 2 & 50.00\\\hline
1st & 19 & \cve{CVE-2009-2560} & 1 & 1 & 50.00\\\hline
1st & 20 & \cve{CVE-2009-3242} & 1 & 1 & 50.00\\\hline
2nd & 1 & \cve{CVE-2009-3829} & 1 & 0 & 100.00\\\hline
2nd & 2 & \cve{CVE-2009-2563} & 1 & 0 & 100.00\\\hline
2nd & 3 & \cve{CVE-2009-2562} & 1 & 0 & 100.00\\\hline
2nd & 4 & \cve{CVE-2009-4378} & 1 & 0 & 100.00\\\hline
2nd & 5 & \cve{CVE-2009-2561} & 1 & 0 & 100.00\\\hline
2nd & 6 & \cve{CVE-2009-4377} & 1 & 0 & 100.00\\\hline
2nd & 7 & \cve{CVE-2009-4376} & 1 & 0 & 100.00\\\hline
2nd & 8 & \cve{CVE-2010-2286} & 1 & 0 & 100.00\\\hline
2nd & 9 & \cve{CVE-2010-0304} & 1 & 0 & 100.00\\\hline
2nd & 10 & \cve{CVE-2010-2285} & 1 & 0 & 100.00\\\hline
2nd & 11 & \cve{CVE-2010-2284} & 1 & 0 & 100.00\\\hline
2nd & 12 & \cve{CVE-2010-2283} & 1 & 0 & 100.00\\\hline
2nd & 13 & \cve{CVE-2009-2559} & 1 & 0 & 100.00\\\hline
2nd & 14 & \cve{CVE-2009-3550} & 1 & 0 & 100.00\\\hline
2nd & 15 & \cve{CVE-2009-3549} & 1 & 0 & 100.00\\\hline
2nd & 16 & \cve{CVE-2010-1455} & 8 & 1 & 88.89\\\hline
2nd & 17 & \cve{CVE-2009-3243} & 3 & 1 & 75.00\\\hline
2nd & 18 & \cve{CVE-2009-3241} & 3 & 1 & 75.00\\\hline
2nd & 19 & \cve{CVE-2009-2560} & 1 & 1 & 50.00\\\hline
2nd & 20 & \cve{CVE-2009-3242} & 1 & 1 & 50.00\\\hline
\end{tabular}

\normalsize
\end{table}

\begin{table}%
\centering
\caption{CVE NLP Stats for \wiresharkCase{0}, Quick Enriched, version SATE.4}
\label{tab:wireshark120-sate4-cwes-quick-enriched}
\begin{tabular}{|c|r|l|c|c|r|}\hline
guess & run & algorithms & good & bad &  \% \\\hline
1st & 1 & \option{-cweid} \option{-nopreprep} \option{-raw} \option{-fft} \option{-cheb}  & 31 & 5 & 86.11\\\hline
1st & 2 & \option{-cweid} \option{-nopreprep} \option{-raw} \option{-fft} \option{-diff}  & 31 & 5 & 86.11\\\hline
1st & 3 & \option{-cweid} \option{-nopreprep} \option{-raw} \option{-fft} \option{-eucl}  & 29 & 7 & 80.56\\\hline
1st & 4 & \option{-cweid} \option{-nopreprep} \option{-raw} \option{-fft} \option{-hamming}  & 22 & 14 & 61.11\\\hline
1st & 5 & \option{-cweid} \option{-nopreprep} \option{-raw} \option{-fft} \option{-cos}  & 33 & 25 & 56.90\\\hline
1st & 6 & \option{-cweid} \option{-nopreprep} \option{-raw} \option{-fft} \option{-mink}  & 20 & 16 & 55.56\\\hline
2nd & 1 & \option{-cweid} \option{-nopreprep} \option{-raw} \option{-fft} \option{-cheb}  & 33 & 3 & 91.67\\\hline
2nd & 2 & \option{-cweid} \option{-nopreprep} \option{-raw} \option{-fft} \option{-diff}  & 33 & 3 & 91.67\\\hline
2nd & 3 & \option{-cweid} \option{-nopreprep} \option{-raw} \option{-fft} \option{-eucl}  & 33 & 3 & 91.67\\\hline
2nd & 4 & \option{-cweid} \option{-nopreprep} \option{-raw} \option{-fft} \option{-hamming}  & 27 & 9 & 75.00\\\hline
2nd & 5 & \option{-cweid} \option{-nopreprep} \option{-raw} \option{-fft} \option{-cos}  & 41 & 17 & 70.69\\\hline
2nd & 6 & \option{-cweid} \option{-nopreprep} \option{-raw} \option{-fft} \option{-mink}  & 22 & 14 & 61.11\\\hline
guess & run & class & good & bad &  \% \\\hline
1st & 1 & \cwe{CWE-399} & 6 & 0 & 100.00\\\hline
1st & 2 & \cwe{NVD-CWE-Other} & 17 & 3 & 85.00\\\hline
1st & 3 & \cwe{CWE-20} & 50 & 10 & 83.33\\\hline
1st & 4 & \cwe{CWE-189} & 8 & 2 & 80.00\\\hline
1st & 5 & \cwe{NVD-CWE-noinfo} & 72 & 40 & 64.29\\\hline
1st & 6 & \cwe{CWE-119} & 13 & 17 & 43.33\\\hline
2nd & 1 & \cwe{CWE-399} & 6 & 0 & 100.00\\\hline
2nd & 2 & \cwe{NVD-CWE-Other} & 17 & 3 & 85.00\\\hline
2nd & 3 & \cwe{CWE-20} & 52 & 8 & 86.67\\\hline
2nd & 4 & \cwe{CWE-189} & 8 & 2 & 80.00\\\hline
2nd & 5 & \cwe{NVD-CWE-noinfo} & 83 & 29 & 74.11\\\hline
2nd & 6 & \cwe{CWE-119} & 23 & 7 & 76.67\\\hline
\end{tabular}

\normalsize
\end{table}

\begin{table}%
\centering
\caption{CVE Stats for \chromeCase{54}, Quick Enriched, (clean CVEs) version SATE.4}
\label{tab:chrome54-sate4-cves-quick-enriched-clean-cvs}
\begin{tabular}{|c|r|l|c|c|r|}\hline
guess & run & algorithms & good & bad &  \% \\\hline
1st & 1 & \option{-nopreprep} \option{-raw} \option{-fft} \option{-eucl}  & 10 & 1 & 90.91\\\hline
1st & 2 & \option{-nopreprep} \option{-raw} \option{-fft} \option{-cos}  & 10 & 1 & 90.91\\\hline
1st & 3 & \option{-nopreprep} \option{-raw} \option{-fft} \option{-diff}  & 10 & 1 & 90.91\\\hline
1st & 4 & \option{-nopreprep} \option{-raw} \option{-fft} \option{-cheb}  & 10 & 1 & 90.91\\\hline
1st & 5 & \option{-nopreprep} \option{-raw} \option{-fft} \option{-mink}  & 9 & 2 & 81.82\\\hline
1st & 6 & \option{-nopreprep} \option{-raw} \option{-fft} \option{-hamming}  & 9 & 2 & 81.82\\\hline
2nd & 1 & \option{-nopreprep} \option{-raw} \option{-fft} \option{-eucl}  & 11 & 0 & 100.00\\\hline
2nd & 2 & \option{-nopreprep} \option{-raw} \option{-fft} \option{-cos}  & 11 & 0 & 100.00\\\hline
2nd & 3 & \option{-nopreprep} \option{-raw} \option{-fft} \option{-diff}  & 11 & 0 & 100.00\\\hline
2nd & 4 & \option{-nopreprep} \option{-raw} \option{-fft} \option{-cheb}  & 11 & 0 & 100.00\\\hline
2nd & 5 & \option{-nopreprep} \option{-raw} \option{-fft} \option{-mink}  & 10 & 1 & 90.91\\\hline
2nd & 6 & \option{-nopreprep} \option{-raw} \option{-fft} \option{-hamming}  & 10 & 1 & 90.91\\\hline
guess & run & class & good & bad &  \% \\\hline
1st & 1 & \cve{CVE-2010-2301} & 6 & 0 & 100.00\\\hline
1st & 2 & \cve{CVE-2010-2300} & 6 & 0 & 100.00\\\hline
1st & 3 & \cve{CVE-2010-2299} & 6 & 0 & 100.00\\\hline
1st & 4 & \cve{CVE-2010-2298} & 6 & 0 & 100.00\\\hline
1st & 5 & \cve{CVE-2010-2297} & 6 & 0 & 100.00\\\hline
1st & 6 & \cve{CVE-2010-2304} & 6 & 0 & 100.00\\\hline
1st & 7 & \cve{CVE-2010-2303} & 6 & 0 & 100.00\\\hline
1st & 8 & \cve{CVE-2010-2295} & 10 & 2 & 83.33\\\hline
1st & 9 & \cve{CVE-2010-2302} & 6 & 6 & 50.00\\\hline
2nd & 1 & \cve{CVE-2010-2301} & 6 & 0 & 100.00\\\hline
2nd & 2 & \cve{CVE-2010-2300} & 6 & 0 & 100.00\\\hline
2nd & 3 & \cve{CVE-2010-2299} & 6 & 0 & 100.00\\\hline
2nd & 4 & \cve{CVE-2010-2298} & 6 & 0 & 100.00\\\hline
2nd & 5 & \cve{CVE-2010-2297} & 6 & 0 & 100.00\\\hline
2nd & 6 & \cve{CVE-2010-2304} & 6 & 0 & 100.00\\\hline
2nd & 7 & \cve{CVE-2010-2303} & 6 & 0 & 100.00\\\hline
2nd & 8 & \cve{CVE-2010-2295} & 10 & 2 & 83.33\\\hline
2nd & 9 & \cve{CVE-2010-2302} & 12 & 0 & 100.00\\\hline
\end{tabular}

\normalsize
\end{table}

\begin{table}%
\centering
\caption{CWE Stats for \chromeCase{54}, (clean CVEs) version SATE.5}
\label{tab:chrome54-sate5-cwes}
\begin{tabular}{|c|r|l|c|c|r|}\hline
guess & run & algorithms & good & bad &  \% \\\hline
1st & 1 & \option{-cweid} \option{-nopreprep} \option{-raw} \option{-fft} \option{-cheb}  & 9 & 0 & 100.00\\\hline
1st & 2 & \option{-cweid} \option{-nopreprep} \option{-raw} \option{-fft} \option{-cos}  & 9 & 0 & 100.00\\\hline
1st & 3 & \option{-cweid} \option{-nopreprep} \option{-raw} \option{-fft} \option{-diff}  & 9 & 0 & 100.00\\\hline
1st & 4 & \option{-cweid} \option{-nopreprep} \option{-raw} \option{-fft} \option{-eucl}  & 8 & 1 & 88.89\\\hline
1st & 5 & \option{-cweid} \option{-nopreprep} \option{-raw} \option{-fft} \option{-hamming}  & 8 & 1 & 88.89\\\hline
1st & 6 & \option{-cweid} \option{-nopreprep} \option{-raw} \option{-fft} \option{-mink}  & 6 & 3 & 66.67\\\hline
2nd & 1 & \option{-cweid} \option{-nopreprep} \option{-raw} \option{-fft} \option{-cheb}  & 9 & 0 & 100.00\\\hline
2nd & 2 & \option{-cweid} \option{-nopreprep} \option{-raw} \option{-fft} \option{-cos}  & 9 & 0 & 100.00\\\hline
2nd & 3 & \option{-cweid} \option{-nopreprep} \option{-raw} \option{-fft} \option{-diff}  & 9 & 0 & 100.00\\\hline
2nd & 4 & \option{-cweid} \option{-nopreprep} \option{-raw} \option{-fft} \option{-eucl}  & 8 & 1 & 88.89\\\hline
2nd & 5 & \option{-cweid} \option{-nopreprep} \option{-raw} \option{-fft} \option{-hamming}  & 8 & 1 & 88.89\\\hline
2nd & 6 & \option{-cweid} \option{-nopreprep} \option{-raw} \option{-fft} \option{-mink}  & 8 & 1 & 88.89\\\hline
guess & run & class & good & bad &  \% \\\hline
1st & 1 & \cwe{CWE-79} & 6 & 0 & 100.00\\\hline
1st & 2 & \cwe{NVD-CWE-noinfo} & 6 & 0 & 100.00\\\hline
1st & 3 & \cwe{CWE-399} & 6 & 0 & 100.00\\\hline
1st & 4 & \cwe{CWE-119} & 6 & 0 & 100.00\\\hline
1st & 5 & \cwe{CWE-20} & 6 & 0 & 100.00\\\hline
1st & 6 & \cwe{NVD-CWE-Other} & 10 & 2 & 83.33\\\hline
1st & 7 & \cwe{CWE-94} & 9 & 3 & 75.00\\\hline
2nd & 1 & \cwe{CWE-79} & 6 & 0 & 100.00\\\hline
2nd & 2 & \cwe{NVD-CWE-noinfo} & 6 & 0 & 100.00\\\hline
2nd & 3 & \cwe{CWE-399} & 6 & 0 & 100.00\\\hline
2nd & 4 & \cwe{CWE-119} & 6 & 0 & 100.00\\\hline
2nd & 5 & \cwe{CWE-20} & 6 & 0 & 100.00\\\hline
2nd & 6 & \cwe{NVD-CWE-Other} & 11 & 1 & 91.67\\\hline
2nd & 7 & \cwe{CWE-94} & 10 & 2 & 83.33\\\hline
\end{tabular}

\normalsize
\end{table}

\begin{table}%
\centering
\caption{CVE Stats for \tomcatCase{13}, version SATE.5}
\label{tab:tomcat13-sate5-cves}
\tiny
\begin{tabular}{|c|r|l|c|c|r|}\hline
1st & 1 & \option{-nopreprep} \option{-raw} \option{-fft} \option{-diff}  & 36 & 7 & 83.72\\\hline
1st & 2 & \option{-nopreprep} \option{-raw} \option{-fft} \option{-cheb}  & 36 & 7 & 83.72\\\hline
1st & 3 & \option{-nopreprep} \option{-raw} \option{-fft} \option{-cos}  & 37 & 9 & 80.43\\\hline
1st & 4 & \option{-nopreprep} \option{-raw} \option{-fft} \option{-eucl}  & 34 & 9 & 79.07\\\hline
1st & 5 & \option{-nopreprep} \option{-raw} \option{-fft} \option{-mink}  & 28 & 15 & 65.12\\\hline
1st & 6 & \option{-nopreprep} \option{-raw} \option{-fft} \option{-hamming}  & 26 & 17 & 60.47\\\hline
2nd & 1 & \option{-nopreprep} \option{-raw} \option{-fft} \option{-diff}  & 40 & 3 & 93.02\\\hline
2nd & 2 & \option{-nopreprep} \option{-raw} \option{-fft} \option{-cheb}  & 40 & 3 & 93.02\\\hline
2nd & 3 & \option{-nopreprep} \option{-raw} \option{-fft} \option{-cos}  & 40 & 6 & 86.96\\\hline
2nd & 4 & \option{-nopreprep} \option{-raw} \option{-fft} \option{-eucl}  & 36 & 7 & 83.72\\\hline
2nd & 5 & \option{-nopreprep} \option{-raw} \option{-fft} \option{-mink}  & 31 & 12 & 72.09\\\hline
2nd & 6 & \option{-nopreprep} \option{-raw} \option{-fft} \option{-hamming}  & 29 & 14 & 67.44\\\hline
guess & run & algorithms & good & bad &  \% \\\hline
1st & 1 & \cve{CVE-2006-7197} & 6 & 0 & 100.00\\\hline
1st & 2 & \cve{CVE-2006-7196} & 6 & 0 & 100.00\\\hline
1st & 3 & \cve{CVE-2006-7195} & 6 & 0 & 100.00\\\hline
1st & 4 & \cve{CVE-2009-0033} & 6 & 0 & 100.00\\\hline
1st & 5 & \cve{CVE-2007-3386} & 6 & 0 & 100.00\\\hline
1st & 6 & \cve{CVE-2009-2901} & 3 & 0 & 100.00\\\hline
1st & 7 & \cve{CVE-2007-3385} & 6 & 0 & 100.00\\\hline
1st & 8 & \cve{CVE-2008-2938} & 6 & 0 & 100.00\\\hline
1st & 9 & \cve{CVE-2007-3382} & 6 & 0 & 100.00\\\hline
1st & 10 & \cve{CVE-2007-5461} & 6 & 0 & 100.00\\\hline
1st & 11 & \cve{CVE-2007-6286} & 6 & 0 & 100.00\\\hline
1st & 12 & \cve{CVE-2007-1858} & 6 & 0 & 100.00\\\hline
1st & 13 & \cve{CVE-2008-0128} & 6 & 0 & 100.00\\\hline
1st & 14 & \cve{CVE-2007-2450} & 6 & 0 & 100.00\\\hline
1st & 15 & \cve{CVE-2009-3548} & 6 & 0 & 100.00\\\hline
1st & 16 & \cve{CVE-2009-0580} & 6 & 0 & 100.00\\\hline
1st & 17 & \cve{CVE-2007-1355} & 6 & 0 & 100.00\\\hline
1st & 18 & \cve{CVE-2008-2370} & 6 & 0 & 100.00\\\hline
1st & 19 & \cve{CVE-2008-4308} & 6 & 0 & 100.00\\\hline
1st & 20 & \cve{CVE-2007-5342} & 6 & 0 & 100.00\\\hline
1st & 21 & \cve{CVE-2008-5515} & 19 & 5 & 79.17\\\hline
1st & 22 & \cve{CVE-2009-0783} & 11 & 4 & 73.33\\\hline
1st & 23 & \cve{CVE-2008-1232} & 13 & 5 & 72.22\\\hline
1st & 24 & \cve{CVE-2008-5519} & 6 & 6 & 50.00\\\hline
1st & 25 & \cve{CVE-2007-5333} & 6 & 6 & 50.00\\\hline
1st & 26 & \cve{CVE-2008-1947} & 6 & 6 & 50.00\\\hline
1st & 27 & \cve{CVE-2009-0781} & 6 & 6 & 50.00\\\hline
1st & 28 & \cve{CVE-2007-0450} & 5 & 7 & 41.67\\\hline
1st & 29 & \cve{CVE-2007-2449} & 6 & 12 & 33.33\\\hline
1st & 30 & \cve{CVE-2009-2693} & 2 & 6 & 25.00\\\hline
1st & 31 & \cve{CVE-2009-2902} & 0 & 1 & 0.00\\\hline
2nd & 1 & \cve{CVE-2006-7197} & 6 & 0 & 100.00\\\hline
2nd & 2 & \cve{CVE-2006-7196} & 6 & 0 & 100.00\\\hline
2nd & 3 & \cve{CVE-2006-7195} & 6 & 0 & 100.00\\\hline
2nd & 4 & \cve{CVE-2009-0033} & 6 & 0 & 100.00\\\hline
2nd & 5 & \cve{CVE-2007-3386} & 6 & 0 & 100.00\\\hline
2nd & 6 & \cve{CVE-2009-2901} & 3 & 0 & 100.00\\\hline
2nd & 7 & \cve{CVE-2007-3385} & 6 & 0 & 100.00\\\hline
2nd & 8 & \cve{CVE-2008-2938} & 6 & 0 & 100.00\\\hline
2nd & 9 & \cve{CVE-2007-3382} & 6 & 0 & 100.00\\\hline
2nd & 10 & \cve{CVE-2007-5461} & 6 & 0 & 100.00\\\hline
2nd & 11 & \cve{CVE-2007-6286} & 6 & 0 & 100.00\\\hline
2nd & 12 & \cve{CVE-2007-1858} & 6 & 0 & 100.00\\\hline
2nd & 13 & \cve{CVE-2008-0128} & 6 & 0 & 100.00\\\hline
2nd & 14 & \cve{CVE-2007-2450} & 6 & 0 & 100.00\\\hline
2nd & 15 & \cve{CVE-2009-3548} & 6 & 0 & 100.00\\\hline
2nd & 16 & \cve{CVE-2009-0580} & 6 & 0 & 100.00\\\hline
2nd & 17 & \cve{CVE-2007-1355} & 6 & 0 & 100.00\\\hline
2nd & 18 & \cve{CVE-2008-2370} & 6 & 0 & 100.00\\\hline
2nd & 19 & \cve{CVE-2008-4308} & 6 & 0 & 100.00\\\hline
2nd & 20 & \cve{CVE-2007-5342} & 6 & 0 & 100.00\\\hline
2nd & 21 & \cve{CVE-2008-5515} & 19 & 5 & 79.17\\\hline
2nd & 22 & \cve{CVE-2009-0783} & 12 & 3 & 80.00\\\hline
2nd & 23 & \cve{CVE-2008-1232} & 13 & 5 & 72.22\\\hline
2nd & 24 & \cve{CVE-2008-5519} & 12 & 0 & 100.00\\\hline
2nd & 25 & \cve{CVE-2007-5333} & 6 & 6 & 50.00\\\hline
2nd & 26 & \cve{CVE-2008-1947} & 6 & 6 & 50.00\\\hline
2nd & 27 & \cve{CVE-2009-0781} & 12 & 0 & 100.00\\\hline
2nd & 28 & \cve{CVE-2007-0450} & 7 & 5 & 58.33\\\hline
2nd & 29 & \cve{CVE-2007-2449} & 8 & 10 & 44.44\\\hline
2nd & 30 & \cve{CVE-2009-2693} & 4 & 4 & 50.00\\\hline
2nd & 31 & \cve{CVE-2009-2902} & 0 & 1 & 0.00\\\hline
\end{tabular}

\normalsize
\end{table}

\begin{table}%
\centering
\caption{CWE Stats for \tomcatCase{13}, version SATE.5}
\label{tab:tomcat13-sate5-cwes}
\begin{tabular}{|c|r|l|c|c|r|}\hline
guess & run & algorithms & good & bad &  \% \\\hline
1st & 1 & \option{-cweid} \option{-nopreprep} \option{-raw} \option{-fft} \option{-cheb}  & 27 & 6 & 81.82\\\hline
1st & 2 & \option{-cweid} \option{-nopreprep} \option{-raw} \option{-fft} \option{-diff}  & 27 & 6 & 81.82\\\hline
1st & 3 & \option{-cweid} \option{-nopreprep} \option{-raw} \option{-fft} \option{-cos}  & 24 & 9 & 72.73\\\hline
1st & 4 & \option{-cweid} \option{-nopreprep} \option{-raw} \option{-fft} \option{-eucl}  & 13 & 20 & 39.39\\\hline
1st & 5 & \option{-cweid} \option{-nopreprep} \option{-raw} \option{-fft} \option{-hamming}  & 12 & 21 & 36.36\\\hline
1st & 6 & \option{-cweid} \option{-nopreprep} \option{-raw} \option{-fft} \option{-mink}  & 9 & 24 & 27.27\\\hline
2nd & 1 & \option{-cweid} \option{-nopreprep} \option{-raw} \option{-fft} \option{-cheb}  & 32 & 1 & 96.97\\\hline
2nd & 2 & \option{-cweid} \option{-nopreprep} \option{-raw} \option{-fft} \option{-diff}  & 32 & 1 & 96.97\\\hline
2nd & 3 & \option{-cweid} \option{-nopreprep} \option{-raw} \option{-fft} \option{-cos}  & 29 & 4 & 87.88\\\hline
2nd & 4 & \option{-cweid} \option{-nopreprep} \option{-raw} \option{-fft} \option{-eucl}  & 17 & 16 & 51.52\\\hline
2nd & 5 & \option{-cweid} \option{-nopreprep} \option{-raw} \option{-fft} \option{-hamming}  & 18 & 15 & 54.55\\\hline
2nd & 6 & \option{-cweid} \option{-nopreprep} \option{-raw} \option{-fft} \option{-mink}  & 13 & 20 & 39.39\\\hline
guess & run & class & good & bad &  \% \\\hline
1st & 1 & \cwe{CWE-264} & 7 & 0 & 100.00\\\hline
1st & 2 & \cwe{CWE-255} & 6 & 0 & 100.00\\\hline
1st & 3 & \cwe{CWE-16} & 6 & 0 & 100.00\\\hline
1st & 4 & \cwe{CWE-119} & 6 & 0 & 100.00\\\hline
1st & 5 & \cwe{CWE-20} & 6 & 0 & 100.00\\\hline
1st & 6 & \cwe{CWE-200} & 22 & 4 & 84.62\\\hline
1st & 7 & \cwe{CWE-79} & 24 & 21 & 53.33\\\hline
1st & 8 & \cwe{CWE-22} & 35 & 61 & 36.46\\\hline
2nd & 1 & \cwe{CWE-264} & 7 & 0 & 100.00\\\hline
2nd & 2 & \cwe{CWE-255} & 6 & 0 & 100.00\\\hline
2nd & 3 & \cwe{CWE-16} & 6 & 0 & 100.00\\\hline
2nd & 4 & \cwe{CWE-119} & 6 & 0 & 100.00\\\hline
2nd & 5 & \cwe{CWE-20} & 6 & 0 & 100.00\\\hline
2nd & 6 & \cwe{CWE-200} & 23 & 3 & 88.46\\\hline
2nd & 7 & \cwe{CWE-79} & 30 & 15 & 66.67\\\hline
2nd & 8 & \cwe{CWE-22} & 57 & 39 & 59.38\\\hline
\end{tabular}

\normalsize
\end{table}

\begin{table}%
\centering
\caption{CVE NLP Stats for \tomcatCase{13}, version SATE.5}
\label{tab:tomcat13-sate5-nlp-cves}
\scriptsize
\begin{tabular}{|c|r|l|c|c|r|}\hline
guess & run & algorithms & good & bad &  \% \\\hline
1st & 1 & \option{-nopreprep} \option{-char} \option{-unigram} -add-delta  & 29 & 4 & 87.88\\\hline
2nd & 1 & \option{-nopreprep} \option{-char} \option{-unigram} -add-delta  & 29 & 4 & 87.88\\\hline
guess & run & class & good & bad &  \% \\\hline
1st & 1 & \cve{CVE-2006-7197} & 1 & 0 & 100.00\\\hline
1st & 2 & \cve{CVE-2006-7196} & 1 & 0 & 100.00\\\hline
1st & 3 & \cve{CVE-2009-2901} & 1 & 0 & 100.00\\\hline
1st & 4 & \cve{CVE-2006-7195} & 1 & 0 & 100.00\\\hline
1st & 5 & \cve{CVE-2009-0033} & 1 & 0 & 100.00\\\hline
1st & 6 & \cve{CVE-2007-1355} & 1 & 0 & 100.00\\\hline
1st & 7 & \cve{CVE-2007-5342} & 1 & 0 & 100.00\\\hline
1st & 8 & \cve{CVE-2009-2693} & 1 & 0 & 100.00\\\hline
1st & 9 & \cve{CVE-2009-0783} & 1 & 0 & 100.00\\\hline
1st & 10 & \cve{CVE-2008-2370} & 1 & 0 & 100.00\\\hline
1st & 11 & \cve{CVE-2007-2450} & 1 & 0 & 100.00\\\hline
1st & 12 & \cve{CVE-2008-2938} & 1 & 0 & 100.00\\\hline
1st & 13 & \cve{CVE-2007-2449} & 3 & 0 & 100.00\\\hline
1st & 14 & \cve{CVE-2007-1858} & 1 & 0 & 100.00\\\hline
1st & 15 & \cve{CVE-2008-4308} & 1 & 0 & 100.00\\\hline
1st & 16 & \cve{CVE-2008-0128} & 1 & 0 & 100.00\\\hline
1st & 17 & \cve{CVE-2009-3548} & 1 & 0 & 100.00\\\hline
1st & 18 & \cve{CVE-2007-5461} & 1 & 0 & 100.00\\\hline
1st & 19 & \cve{CVE-2007-3382} & 1 & 0 & 100.00\\\hline
1st & 20 & \cve{CVE-2007-0450} & 2 & 0 & 100.00\\\hline
1st & 21 & \cve{CVE-2009-0580} & 1 & 0 & 100.00\\\hline
1st & 22 & \cve{CVE-2007-6286} & 1 & 0 & 100.00\\\hline
1st & 23 & \cve{CVE-2008-5515} & 3 & 1 & 75.00\\\hline
1st & 24 & \cve{CVE-2008-1232} & 1 & 2 & 33.33\\\hline
1st & 25 & \cve{CVE-2009-2902} & 0 & 1 & 0.00\\\hline
2nd & 1 & \cve{CVE-2006-7197} & 1 & 0 & 100.00\\\hline
2nd & 2 & \cve{CVE-2006-7196} & 1 & 0 & 100.00\\\hline
2nd & 3 & \cve{CVE-2009-2901} & 1 & 0 & 100.00\\\hline
2nd & 4 & \cve{CVE-2006-7195} & 1 & 0 & 100.00\\\hline
2nd & 5 & \cve{CVE-2009-0033} & 1 & 0 & 100.00\\\hline
2nd & 6 & \cve{CVE-2007-1355} & 1 & 0 & 100.00\\\hline
2nd & 7 & \cve{CVE-2007-5342} & 1 & 0 & 100.00\\\hline
2nd & 8 & \cve{CVE-2009-2693} & 1 & 0 & 100.00\\\hline
2nd & 9 & \cve{CVE-2009-0783} & 1 & 0 & 100.00\\\hline
2nd & 10 & \cve{CVE-2008-2370} & 1 & 0 & 100.00\\\hline
2nd & 11 & \cve{CVE-2007-2450} & 1 & 0 & 100.00\\\hline
2nd & 12 & \cve{CVE-2008-2938} & 1 & 0 & 100.00\\\hline
2nd & 13 & \cve{CVE-2007-2449} & 3 & 0 & 100.00\\\hline
2nd & 14 & \cve{CVE-2007-1858} & 1 & 0 & 100.00\\\hline
2nd & 15 & \cve{CVE-2008-4308} & 1 & 0 & 100.00\\\hline
2nd & 16 & \cve{CVE-2008-0128} & 1 & 0 & 100.00\\\hline
2nd & 17 & \cve{CVE-2009-3548} & 1 & 0 & 100.00\\\hline
2nd & 18 & \cve{CVE-2007-5461} & 1 & 0 & 100.00\\\hline
2nd & 19 & \cve{CVE-2007-3382} & 1 & 0 & 100.00\\\hline
2nd & 20 & \cve{CVE-2007-0450} & 2 & 0 & 100.00\\\hline
2nd & 21 & \cve{CVE-2009-0580} & 1 & 0 & 100.00\\\hline
2nd & 22 & \cve{CVE-2007-6286} & 1 & 0 & 100.00\\\hline
2nd & 23 & \cve{CVE-2008-5515} & 3 & 1 & 75.00\\\hline
2nd & 24 & \cve{CVE-2008-1232} & 1 & 2 & 33.33\\\hline
2nd & 25 & \cve{CVE-2009-2902} & 0 & 1 & 0.00\\\hline
\end{tabular}

\normalsize
\end{table}

\begin{table}%
\centering
\caption{CWE NLP Stats for \tomcatCase{13}, version SATE.5}
\label{tab:tomcat13-sate5-nlp-cwes}
\NIST{}
\begin{tabular}{|c|r|l|c|c|r|}\hline
guess & run & algorithms & good & bad &  \% \\\hline
1st & 1 & \option{-cweid} \option{-nopreprep} \option{-char} \option{-unigram} -add-delta  & 13 & 20 & 39.39\\\hline
2nd & 1 & \option{-cweid} \option{-nopreprep} \option{-char} \option{-unigram} -add-delta  & 17 & 16 & 51.52\\\hline
guess & run & class & good & bad &  \% \\\hline
1st & 1 & \cwe{CWE-16} & 1 & 0 & 100.00\\\hline
1st & 2 & \cwe{CWE-255} & 1 & 0 & 100.00\\\hline
1st & 3 & \cwe{CWE-264} & 2 & 0 & 100.00\\\hline
1st & 4 & \cwe{CWE-119} & 1 & 0 & 100.00\\\hline
1st & 5 & \cwe{CWE-20} & 1 & 0 & 100.00\\\hline
1st & 6 & \cwe{CWE-200} & 3 & 1 & 75.00\\\hline
1st & 7 & \cwe{CWE-22} & 3 & 13 & 18.75\\\hline
1st & 8 & \cwe{CWE-79} & 1 & 6 & 14.29\\\hline
2nd & 1 & \cwe{CWE-16} & 1 & 0 & 100.00\\\hline
2nd & 2 & \cwe{CWE-255} & 1 & 0 & 100.00\\\hline
2nd & 3 & \cwe{CWE-264} & 2 & 0 & 100.00\\\hline
2nd & 4 & \cwe{CWE-119} & 1 & 0 & 100.00\\\hline
2nd & 5 & \cwe{CWE-20} & 1 & 0 & 100.00\\\hline
2nd & 6 & \cwe{CWE-200} & 4 & 0 & 100.00\\\hline
2nd & 7 & \cwe{CWE-22} & 5 & 11 & 31.25\\\hline
2nd & 8 & \cwe{CWE-79} & 2 & 5 & 28.57\\\hline
\end{tabular}

\normalsize
\end{table}

\begin{table}%
\centering
\caption{CVE NLP Stats for \chromeCase{54}, version SATE.7}
\label{tab:chrome54-sate7-nlp-cves}
\NIST{}
\begin{tabular}{|c|r|l|c|c|r|}\hline
guess & run & algorithms & good & bad &  \% \\\hline
1st & 1 & \option{-nopreprep} \option{-char} \option{-unigram} -add-delta  & 9 & 0 & 100.00\\\hline
2nd & 1 & \option{-nopreprep} \option{-char} \option{-unigram} -add-delta  & 9 & 0 & 100.00\\\hline
guess & run & class & good & bad &  \% \\\hline
1st & 1 & \cve{CVE-2010-2304} & 1 & 0 & 100.00\\\hline
1st & 2 & \cve{CVE-2010-2298} & 1 & 0 & 100.00\\\hline
1st & 3 & \cve{CVE-2010-2301} & 1 & 0 & 100.00\\\hline
1st & 4 & \cve{CVE-2010-2295} & 2 & 0 & 100.00\\\hline
1st & 5 & \cve{CVE-2010-2300} & 1 & 0 & 100.00\\\hline
1st & 6 & \cve{CVE-2010-2303} & 1 & 0 & 100.00\\\hline
1st & 7 & \cve{CVE-2010-2297} & 1 & 0 & 100.00\\\hline
1st & 8 & \cve{CVE-2010-2299} & 1 & 0 & 100.00\\\hline
2nd & 1 & \cve{CVE-2010-2304} & 1 & 0 & 100.00\\\hline
2nd & 2 & \cve{CVE-2010-2298} & 1 & 0 & 100.00\\\hline
2nd & 3 & \cve{CVE-2010-2301} & 1 & 0 & 100.00\\\hline
2nd & 4 & \cve{CVE-2010-2295} & 2 & 0 & 100.00\\\hline
2nd & 5 & \cve{CVE-2010-2300} & 1 & 0 & 100.00\\\hline
2nd & 6 & \cve{CVE-2010-2303} & 1 & 0 & 100.00\\\hline
2nd & 7 & \cve{CVE-2010-2297} & 1 & 0 & 100.00\\\hline
2nd & 8 & \cve{CVE-2010-2299} & 1 & 0 & 100.00\\\hline
\end{tabular}

\normalsize
\end{table}

\begin{table}%
\centering
\caption{CWE NLP Stats for \chromeCase{54}, version SATE.7}
\label{tab:chrome54-sate7-nlp-cwes}
\NIST{}
\begin{tabular}{|c|r|l|c|c|r|}\hline
guess & run & algorithms & good & bad &  \% \\\hline
1st & 1 & \option{-cweid} \option{-nopreprep} \option{-char} \option{-unigram} -add-delta  & 8 & 1 & 88.89\\\hline
2nd & 1 & \option{-cweid} \option{-nopreprep} \option{-char} \option{-unigram} -add-delta  & 8 & 1 & 88.89\\\hline
guess & run & class & good & bad &  \% \\\hline
1st & 1 & \cwe{CWE-399} & 1 & 0 & 100.00\\\hline
1st & 2 & \cwe{NVD-CWE-noinfo} & 1 & 0 & 100.00\\\hline
1st & 3 & \cwe{CWE-79} & 1 & 0 & 100.00\\\hline
1st & 4 & \cwe{NVD-CWE-Other} & 2 & 0 & 100.00\\\hline
1st & 5 & \cwe{CWE-119} & 1 & 0 & 100.00\\\hline
1st & 6 & \cwe{CWE-20} & 1 & 0 & 100.00\\\hline
1st & 7 & \cwe{CWE-94} & 1 & 1 & 50.00\\\hline
2nd & 1 & \cwe{CWE-399} & 1 & 0 & 100.00\\\hline
2nd & 2 & \cwe{NVD-CWE-noinfo} & 1 & 0 & 100.00\\\hline
2nd & 3 & \cwe{CWE-79} & 1 & 0 & 100.00\\\hline
2nd & 4 & \cwe{NVD-CWE-Other} & 2 & 0 & 100.00\\\hline
2nd & 5 & \cwe{CWE-119} & 1 & 0 & 100.00\\\hline
2nd & 6 & \cwe{CWE-20} & 1 & 0 & 100.00\\\hline
2nd & 7 & \cwe{CWE-94} & 1 & 1 & 50.00\\\hline
\end{tabular}

\normalsize
\end{table}

\printindex

\end{document}